\documentclass[preprint,showpacs,preprintnumbers,amsmath,amssymb]{revtex4}
\usepackage{epsfig}

\usepackage{graphicx}
\usepackage{dcolumn}
\usepackage{bm}
\def\lsim{\mathrel {\vcenter {\baselineskip 0pt \kern 0pt
    \hbox{$&lt;$} \kern 0pt \hbox{$\sim$} }}}
\def\gsim{\mathrel {\vcenter {\baselineskip 0pt \kern 0pt
    \hbox{$&gt;$} \kern 0pt \hbox{$\sim$} }}}

\newcommand{\U}{{\cal {U}}}

\begin{document}

\title{Unparticle Induced Baryon Number Violating Nucleon Decays}

\author{$^1$Xiao-Gang He}
\email{hexg@phys.ntu.edu.tw}
\author{$^2$Sandip Pakvasa}
\email{pakvasa@physics.hawaii.edu}
\affiliation{ $^1$Department of Physics
and Center for Theoretical Sciences, National Taiwan University,
Taipei, Taiwan\\
Department of Physics and Astronomy, University of Hawaii at Manoa, Hawaii, 96822, USA}

\date{\today}

\begin{abstract}
We study baryon number violating nucleon decays induced by
unparticle interactions with the standard model particles. We find
that the lowest dimension operators which cause nucleon decays can
arise at dimension $6 + (d_s-3/2)$ with the unparticles being a
spinor of dimension $d_s=d_\U +1/2$. For scalar and vector unparticles of dimension $d_\U$,
the lowest order
operatoers arise at $6+d_\U$ and $7+d_\U$ dimensions,
respectively. Comparing the spinor unparticle induced $n \to
O^s_\U$ and experimental bound on invisible decay of a neutron
from KamLAND, we find that the scale for unparticle physics is
required to be larger than $10^{10}$ GeV for $d_\U < 2$ if the couplings
are set to be of order one. For scalar and vector
unparticles, the dominant baryon number violating decay modes are
$n\to \bar \nu + O_\U (O^\mu_\U)$ and $p \to e^+ + O_\U
(O^\mu_\U)$. The same experimental bound puts the scales for
scalar and vector unparticle to be larger than $10^{7}$ and
$10^{5}$ GeV for $d_\U <2$ with couplings set to be of order one.
Data on $p \to e^+ \mbox{invisible}$ puts similar constraints on
unparticle interactions.

\end{abstract}


\maketitle

One of the outstanding problems of modern particle physics it
whether proton is stable or not. If proton decays, baryon number is violated.
Baryon number violation is one of the necessary conditions to explain why our
universe at present is dominated by matter if initially there are
equal amount of matter and anti-matter as shown by Sakharov\cite{sakharov}. It also provides a test
for grand unified theories. Experimentally, no proton decay has
been detected setting a lower bound of
$10^{33}$ years\cite{pdg} for proton lifetime. The Standard Model (SM) Lagrangian
does not allow baryon number violating renormalizable interactions
and therefore forbids proton or more generally baryon number
violating nucleon decays. Although non-perturbative effects can
induce baryon number violation in the SM, it is too small for any
experimental observation. In grand unification theories, such as
$SU(5)$ theory, baryon number can be violated and proton can decay
by exchanging heavy particles. The long lifetime bound on proton
pushes the scale of the heavy particle mass to the unification
scale and rules out some grand unification models. If
nonrenormalizable terms are allowed in the Lagrangian,
it is possible to have baryon number violation in the SM.
The lowest dimension operators of this kind have dimension six. When
going beyond the SM, it is also possible to have renormalizable baryon
number violating interactions at low energy, such as in R-parity violating
supersymmetric theories. It is of interest to determine the
constraints on the interaction strengthes of these
operators. Such studies can provide information about the scale
where the new physics effects become important. Much work has been
done along this line. In this work we study possible effect of
unparticle physics on baryon number violating nucleon decays.

The concept of unparticle \cite{Georgi:2007ek} stems from the
observation that certain high energy theory with a nontrivial
infrared fixed-point at some scale $\Lambda_{\U}$ may develop a
scale-invariant degree of freedom below the scale. The unparticle kinematics
is mainly determined by its scaling dimension $d_{\U}$ under scale
transformations. The unparticle must interact with particles,
however feebly, to be physically relevant; and the interaction can
be well described in effective field theory.
At low energy the interaction of an unparticle $O_\U$ with an operator
composed of SM particles $O_{SM}$ of dimension $d_{SM}$ can be
parameterized in the form $\lambda \Lambda_{\cal{U}}
^{4-d_{SM} - d_\U} O_{SM} O_{\cal{U}}$. There has been a burst of activities on
unparticle studies\cite{unparticle,lou,he,0705,0706,0707,0708,0709,0710,0711} since the
seminal work of Georgi~\cite{Georgi:2007ek}.

An unparticle looks like a non-integral $d_\U$ dimension invisible
particle. Depending on the nature of the original operator
inducing the unparticle and the mechanism of transmutation, the
resulting unparticles may have different Lorentz structures, such
as scalar ($O_{\U}$), spinor ($O_{\U}^s$) and vector
($O^\mu_{U}$) unparticles. We use $d_\U$ to indicate the dimensions of
$O_\U$ and $O_\U^\mu$, and $d_s = d_\U + 1/2$ to indicate the dimension of $O^s_\U$.
When taking the limit that $d_\U = 1$, the operators go to the limit of ordinary scalar,
vector and spinor fields.
There are many unknowns when writing
down the effective interaction with unparticles even if one
assumes that it is a SM singlet with known spin
structure\cite{he}. Most of the phenomenological studies then
focused on constraining unparticle interactions with the SM
particles using various processes. In our study of unparticle
interaction induced baryon number violating nucleon decays we
will also use the effective field theory approach. We first identify
all possible low dimension operators relevant and then constrain
the couplings.

In the SM, operators which can induce baryon number violating
nucleon decays can only be generated at dimension six or higher.
With unparticles, the lowest dimension operators can arise at $6
+ (d_s - 3/2)$ with the unparticles being a spinor. For scalar and vector
unparticles, the lowest order operatoers arise at $6+d_\U$ and
$7+d_\U$ dimensions, respectively. We find that the recent result
on invisible decay of a neutron from KamLAND\cite{kamland} can put very
stringent bounds on the relevant coupling. A reliable bound on proton decay
into a positron and missing energy can also put stringent
constraints. We now proceed with details.

Let us begin by listing the dimension six operators which violate baryon number in the SM,
\begin{eqnarray}
&&O_{QQQ}=\bar Q_L^c Q_L \bar L_L^cQ_L,\;\;O_{QQU}=\bar Q_L^c Q_L \bar E_R^c  U_R,\;\;
O_{DUQ}=\bar D_R^c  U_R \bar L_L^c Q_L,\nonumber\\
&&O_{UUD}=\bar U_R^c U_R\bar E_R^c D_R,\;\;
O_{DUU}=\bar D_R^c U_R \bar E_R^c U_R,\;\;O_{QQD}=\bar Q_L^c Q_L\bar \nu_R^c D_R,\nonumber\\
&&O_{DDU}=\bar D_R^c D_R\bar \nu_R^c U_R, \;\;O_{UDD}=\bar U_R^c D_R \bar \nu_R^c D_R. \label{smo}
\end{eqnarray}
Here we have also included operators involving
right-handed neutrinos which may be needed for neutrino mass in
the Standard Model. Each operator is associated with a coupling
strength $\lambda_i/\Lambda^2$. Here $\Lambda$ is the scale where
the baryon number violation is generated due to new physics
effects. With a given $\Lambda$, $\lambda_i$ indicates the
relative strength of each operator.

The lowest dimension operators which violate baryon number can be constructed involve
spinor unparticles. They are given by
\begin{eqnarray}
O^s_{QQD}=\bar Q_L^cQ_L \bar O_\U^s D_R,\;\;O^s_{UUD}=\bar U_R^c U_R \bar O_\U^s D_R,
\;\;O^s_{DUU}=\bar D_R^c U_R \bar O_\U^s U_R.
\end{eqnarray}
Each operator is associated with a coupling ${\lambda^s_i/
\Lambda_\U^{d_\U+1/2}}$. These operators look similar to the one
with the spinor unparticle replaced by a right-handed singlet
neutrino in form. However, there is a crucial difference that with
unparticle, one can talk about a baryon decay into an unparticle,
but not decay into another particle. We will come back to this
later.

For scalar unparticles, in order to have baryon number violation,
one has to go to at least dimension $6+d_\U$. The lowest dimension ones can be obtained
by attaching a scalar unparticle $O_\U$ to the operators in
eq.(\ref{smo}), such as $O_{QQQ}^\U=\bar Q_L^c Q_L \bar L^c_L Q_L
O_\U$, with $\lambda_i/\Lambda^2$ replaced by
$\lambda^\U_i/\Lambda_\U^{d_\U + 2}$.

For vector unparticles, one has to go to even higher order, at least $7+d_\U$.
The lowest dimension ones can be obtained by attaching the unparticle to the operators
in eq.(\ref{smo}) and
inserting $O_\U^\mu$, derivative $\partial^\mu$ and the
covariant derivative $D_\mu$ in between the bi-spinors in eq.(\ref{smo}) at appropriate places, for example
$\bar Q_L^c Q_L \bar L_L^c \sigma_{\mu\nu}Q_L \partial^\mu O^\nu_\U$. The associated coupling
should then be replaced by
$\lambda^\mu_i/\Lambda_\U^{d_\U + 3}$.

The above operators can induce baryon number violating nucleon
decays. Upper bound on relevant decay modes can
be used to put constraints on the corresponding parameters. Since the
unparticle behaves like an invisible object which carries away
energy and escape detection, the signature is missing energy, the invisible part of the decay. We now
study the constraints on the couplings of the above mentioned
operators.

The baryon number violating operators with a spinor unparticle
will induce $n\to O^s_\U$ decay. The experimental signature is
total invisible decay of a neutron. For this decay there is a
strong recent bound from Kamland\cite{kamland} with $\tau(n \to
\mbox{invisible}) > 5.8\times 10^{30}$ years. Using this bound we
can put a very stringent bound on the couplings. Let us take the
operator $\bar Q_L^cQ_L \bar O^s D_R$ to show details.

The matrix element for this decay is given by
\begin{eqnarray}
M(n\to \U) = 2{\lambda_{QQD}\over \Lambda^{d_\U+1/2}_\U}\alpha \bar O_s Rn,
\end{eqnarray}
where the parameter $\alpha$ is defined by $\alpha R n
=- <0|(\bar u^c_{\beta} L d_{\gamma})R d_{\alpha}\epsilon^{\alpha\beta\gamma}|n>$.
Here $R(L) = (1+(-)\gamma_5)/2$.

Several other related matrix elements will be used later. We summarize the definitions here. They are
$\alpha L n = <0|(\bar u^c_{\beta} R d_{\gamma})L d_{\alpha}\epsilon^{\alpha\beta\gamma}|n>$,
$\beta L n = <0|(\bar u^c_{\beta} L d_{\gamma})L d_{\alpha}\epsilon^{\alpha\beta\gamma}|n>$ and
$\beta R n = -
<0|(\bar u^c_{\beta} R d_{\gamma})R d_{\alpha}\epsilon^{\alpha\beta\gamma}|n>$.
The absolute values of $\alpha$ and $\beta$ are almost equal to each other. These matrix elements have
been calculated on the lattice recently. Calculations in ref.\cite{lattice} give
$\alpha = -0.0118(21)$ GeV$^3$ and $\beta = 0.0118(21)$ GeV$^3$, and calculations in ref.\cite{lattice1} give
$|\alpha| = 0.0090 (+5 - 19)$ GeV$^3$ and $|\beta| = 0.0096(09)(+6 -20)$ GeV$^3$.
In our later discussions we will use 0.01 GeV$^3$ for both $|\alpha|$ and $|\beta|$.

In calculating decay width, one should be careful with the phase space of a unparticle which
is dramatically different than that for a particle.
For a usual massless particle, the phase space is given by $2\pi \theta(p^0)\delta(p^2)d^4p/(2\pi)^4$, but for
a unparticle it is replaced by $A_{d_\U} \theta(p^0)\theta(p^2) a d^4p/(2\pi)^4$. Here
$A_{du} =(16 \pi^{5/2}/(2\pi)^{2d_\U})\Gamma(d_\U+1/2)/(\Gamma(d_\U - 1)\Gamma(2d_\U)$.
For scalar and vector unparticles,
$a = (p^2)^{d_\U-2}$, and for spinor unparticle, $a = (p^2)^{d_s-5/2} = (p^2)^{d_\U - 4}$.

Due to the unique phase structure of
unparticles, a particle of any mass can decay into an unparticle. For $n\to O^s_\U$, we obtain
\begin{eqnarray}
\Gamma(n\to O^s_\U) = 4 A_{d_\U}|\lambda^s_{QQD}|^2{|\alpha|^2\over m^5_n}\left ({m_n\over \Lambda_\U}\right )^{2d_\U+2}.
\label{decay}
\end{eqnarray}
Here we have used parity conserving spin-sum of spinor unparticle field\cite{lou},
$\sum_{spin} O^s_\U  \bar O^s_\U = \gamma_\mu p^\mu$, and $d_s = d_\U + 1/2$.

We comment that in the limit of $d_\U$ equal to 1, the above decay width
becomes zero since $A_\U$ has a factor $1/\Gamma(d_\U -1)$ which goes to zero when $d_\U \to 1$.
Physically this is because that in this limit $A_{d_\U} \theta(p^2)/p^{2(2-d_\U)} \to 2\pi \delta(p^2)$
and the unparticle behaves as a massless particle. When $p^2 = m_n^2$, the delta function forces
the width to be zero.

Saturating the experimental bound on $n\to \mbox{invisible}$ by
the above decay, one can constrain the unparticle interactions. In
Fig.1, we show constraint on $\lambda^s_{QQD}$ as a function of
$d_\U$ for fixed $\Lambda_\U = 10$ TeV.  We see that the
constraint is very stringent. If $\Lambda_\U$ is set to be larger,
the coupling becomes larger.  In Fig. 2, we show the bound on the
scalar $\Lambda_\U$ with $\lambda_i = 1$. Setting $\lambda_{QQD}$
to be of order 1, the unparticle scale $\Lambda_\U$ would be
required to be larger than $10^{10}$ GeV (for $d_\U = 1.5$,
$\Lambda_\U^s$ is around $10^{12}$ GeV).

\begin{figure*}[t!]
\includegraphics[width=4 in]{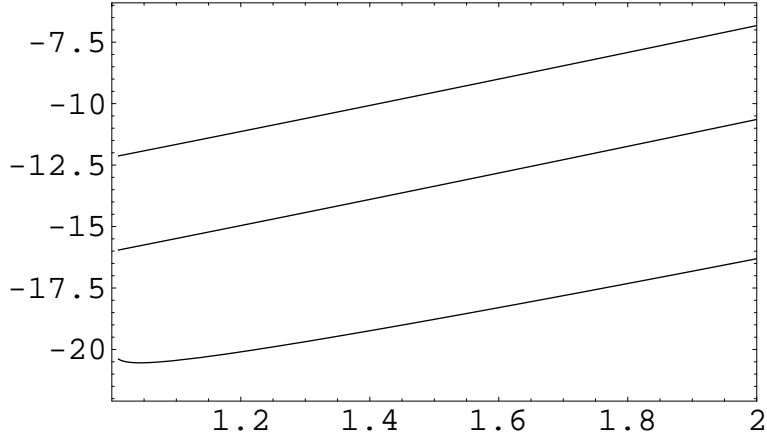}
\caption{\label{f1} \small Constraints on $log_{10}\lambda_i$ as functions of $d_\U$ for $\Lambda_\U = 10$ TeV and
$|\alpha| = |\beta| = 0.01$ GeV$^3$. The curves from bottom up are the upper bounds for $\lambda_{QQD}^s$,
$\lambda^\U_{QQQ}$ and $\lambda^\mu$ from
the processes $n \to + O^s_\U$,
$n\to \bar \nu + O_\U$, and
$n\to \bar \nu + O^\mu_\U$, respectively.}
\end{figure*}

\begin{figure*}[t!]
\includegraphics[width=4 in]{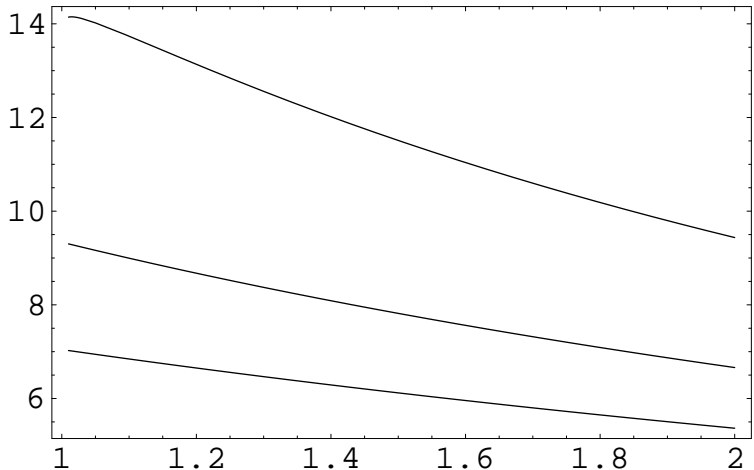}
\caption{\label{f2} \small Constraints on $log_{10}(\Lambda_\U/GeV)$ as
functions of $d_\U$ for $\lambda_i$ and $|\alpha| = |\beta| =
0.01$ GeV$^3$. The curves from top down are the lower bounds for
$log_{10}(\Lambda_\U/GeV)$, from the processes $n \to O^s_\U$, $n\to \bar \nu + O_\U$, and $n\to \bar
\nu + O^\mu_\U$, respectively.}
\end{figure*}

Replacing $2\lambda_{QQD}^s$ by $\lambda_{UDD}^s$ and $\lambda_{DDU}^s$ in eq. (\ref{decay}),
one obtains the decay widths induced by the operators $\bar U^c_R D_R \bar O^s D_R$
and $\bar D_R^c D_R \bar O^s U_R $.

The operators in eq.(\ref{smo}) can induce $p\to e^+ + \pi^0, \bar
\nu + \pi^+$ and $n\to e^+ + \pi^-, \bar \nu + \pi^0$ decays which
have been studied. The couplings are stringently constrained.
When attaching an scalar unparticle $O_\U$ to the operators in eq.(\ref{smo}),
one would naturally consider $p\to e^+ + \pi^0 +
O_\U, \bar \nu + \pi^+ + O_\U$ and $n\to e^+ + \pi^- + O_\U, \bar
\nu + \pi^0 + O_\U$ decay modes to constrain the interactions. We
find, however,  that there are simpler decay modes such as $p\to e^+ +
O_\U$, and $n\to \bar \nu + O_\U$ which can be used to constrain the
interactions. Let us take $\bar Q^c_L Q_L \bar L^c_L Q_L O_\U$ to
show the details. We have the effective Lagrangian for these
decays
\begin{eqnarray}
L(p \to e^+ + O_\U) = 2{\lambda_{QQQ}^\U\over \Lambda_\U^{d_\U + 2}} \beta \bar e^c_L p O_\U,\;\;
L(n \to \bar \nu + O_\U) = - 2{\lambda_{QQQ}^\U\over \Lambda_\U^{d_\U + 2}} \beta \bar \nu ^c_L n O_\U.
\end{eqnarray}
When neutrino and electron masses are neglected, the formulas for the decay width for
$p \to e^+ + O_\U$ and $n\to \bar \nu + O_\U$ are the same. We have
\begin{eqnarray}
\Gamma = A_{d_\U} {|\lambda^\U_{QQQ}|^2\over 16 \pi^2} {|\beta|^2\over m_N \Lambda_\U^4}
\left ({m_N\over \Lambda_\U}\right )^{2d_\U}
B(3,d_\U -1),
\end{eqnarray}
where $m_N = m_p$ and $M_N = m_n$ for proton and neutron decays. $B(a,b)$ is the standard $\beta$-function.

The experimental bound on $n\to \mbox{invisible}$ from Kamland
can be used to constrain $n\to \bar \nu + O_\U$ since neutrino is not
measured. We show the constraint on the parameter $\lambda^\U_{QQQ}$ in Fig. 1.
It can be seen that the constraint is also very strong although weaker than that for $\lambda^s_{QQD}$.
Setting $\lambda^\U_{QQQ}$ to be one, with $d_\U = 1.5$, the scale $\Lambda_\U$
is required to be larger than $10^8$ GeV. More details are shown in Fig. 2.

The experimental signature for the decay mode $p \to e^+ + O_\U$ is $p \to e^+ + \mbox{invisible}$.
If one takes the bound $\tau > 6\times 10^{29}$
years for $p\to e^+ + \mbox{anything}$ from PDG\cite{pdg,jhon} and saturate it with $p\to e^+ + O_\U$,
one would obtain similar constraints as that from $n\to \bar \nu + O_\U$.
The other operators will induce similar decays, and the constraints are also similar.

Finally let us discuss vector unparticle induced baryon number
violating decays. There are many operators at dimension $7+d_\U$
which can induce such decays. For illustration we provide details for
the operator $\bar Q^c_L Q_L \bar L^c_L \sigma_{\mu\nu} Q_L
\partial^\mu O^\nu_\U$. This operator will induce $p\to e^+ +
O_\U^\mu$, and $ n\to \bar \nu  + O_\U^\mu$. The effective Lagrangian for
these decays are given by
\begin{eqnarray}
&&L(p \to e^+ + O_\U) = 2{\lambda_{QQQ}^\mu\over \Lambda_\U^{d_\U + 3}}
\beta \bar e^c_L \sigma_{\mu\nu}p \partial^\mu O^\nu_\U,\nonumber\\
&&L(n \to \bar \nu + O_\U) = - 2{\lambda_{QQQ}^\mu \over \Lambda_\U^{d_\U + 3}}
\beta \bar \nu ^c_L \sigma_{\mu\nu}n \partial^\mu O^\nu_\U.
\end{eqnarray}
Neglecting neutrino and electron masses, we obtain the decay rates for these decays
\begin{eqnarray}
\Gamma = A_{d_\U} {|\lambda^\mu_{QQQ}|^2\over 32 \pi^2}{|\beta|^2\over \Lambda_\U^6}
m_N \left ( {m_N\over \Lambda_\U}\right )^{2d_\U} {6d_\U+9\over d_\U+2}B(3,d-1).
\end{eqnarray}

Again using $n\to \mbox{invisible}$ data, one can constrain the couplings. We show the results in
Fig.1. From the figure it is clear that the constraint is weaker compared with previous constraints,
but is still very strong. Setting $\lambda^\mu_{QQQ}$ to be one, with $d_\U = 1.5$, the scale $\Lambda_\U$
is required to be larger than $10^6$ GeV as can be seen from Fig. 2.

In summary, we have studied baryon number violating nucleon decays
induced by unparticle interactions with the standard model
particles. We found that the lowest dimension operators can arise
at dimension $6+(d_s -3/2) = 6 + (d_\U-1)$ with the unparticles being a spinor. For
scalar and vector unparticles, the lowest order operatoers arise
at $6+d_\U$ and $7+d_\U$ dimensions, respectively. For spinor
unparticle, the dominant decay mode is $n \to O^s_\U$.
Experimental bound on invisible decay of a neutron from KamLAND
puts very stringent bounds on the relevant coupling. If the
coupling is of order one, the unparticle scale is required to be
larger than $10^{10}$ GeV for $d_\U <2$. For scalar and vector unparticles, the
dominant decay modes are $n\to \bar \nu + O_\U (O^\mu_\U)$ and $p
\to e^+ + O_\U (O^\mu_\U)$. Invisible decay of a neutron bound
also puts very strong constraints on the relevant couplings and
push the unparticle scales for sclar and vector to be $10^{7}$ GeV
and $10^5$ GeV for $d_\U <2$, repectively. Data on proton decay into a positron
and missing energy can also put stringent constrains.

\vspace*{0.2cm}
{\bf Acknowledgement}: XGH was supported in part by the NSC and NCTS,
and SP was supported by U.S.D.O.E. under grant no. DE-FG03-91ER40833.
SP would like to acknowledge support and hospitality of the CosPA 2007
(November 2007) where this work was initiated.


\begin{references}
\bibitem{jhon} Although stronger bounds on the decay mode $p\to e^+ + \mbox{invisible}$
can, most likely,  be derived from the existing data (J.G. Learned, Private Communication).
\end{references}
\end{document}